\documentclass[a4paper]{article}

\usepackage[dvipdfmx]{graphicx}
\usepackage{type1cm}
\usepackage{slashbox}
\usepackage{multirow}
\usepackage{lscape}
\usepackage{amsmath}
\usepackage{framed}
\usepackage{setspace}

\usepackage{mathtools}
\DeclarePairedDelimiter{\ceil}{\lceil}{\rceil}

\begin{document}
\fontsize{10.5pt}{18pt}\selectfont



\fontsize{13pt}{18pt}\selectfont
\begin{center}
{\bf Analyses of Aggregate Fluctuations of Firm Network Based on the Self-Organized Criticality Model\footnote{
This study is conducted as a part of the project
``Price Network and Dynamics of Small and Medium Enterprises''
undertaken at the Research Institute of Economy, Trade and Industry (RIETI).
The authors thank the institute for various means of support.
We thank Hiroshi Yoshikawa, Hideaki Aoyama, Hiroshi Iyetomi, Yuichi Ikeda, Yoshi Fujiwara, Wataru Soma, Yoshiyuki Arata,
and members who attended the internal seminar of RIETI for their helpful comments.
We gratefully acknowledge financial support from the Japan Society for the Promotion of Science (No. 15K01217).
}}

\vspace{2ex}

Hiroyasu Inoue\footnote{Graduate School of Simulation Studies, University of Hyogo, 7-1-28 Minatojima-minamimachi, Chuo-ku, Kobe, Hyogo 650-0047, Japan}\\
Graduate School of Simulation Studies, University of Hyogo

\vspace{3ex}

\fontsize{10.5pt}{18pt}\selectfont
Abstract

\end{center}

\fontsize{10.5pt}{18pt}\selectfont
\noindent
This study examine the difference in the size of avalanches among industries triggered by demand shocks, which can be rephrased by control of the economy or fiscal policy, and by using the production-inventory model and observed data. We obtain the following results. (1) The size of avalanches follows power law. (2) The mean sizes of avalanches for industries are diverse but their standard deviations highly overlap. (3) We compare the simulation with an input-output table and with the actual policies. They are compatible. 

\vspace{2ex}


\vspace{5ex}




\vspace{5ex}



\fontsize{10.5pt}{18pt}\selectfont

\newpage

\section{Introduction}

Giving stimulus to firms and prompting the spillover effect is a way for a government to affect its economy,
which includes purchasing goods and services, giving grants to firms, and fine-tuning taxes.
Governments consider fiscal policy as an important determinant of growth \cite{Easterly93,Romp07}.
Currently, analysis of input-output table is considered a strong tool to predict the spillover effect \cite{Leontief36}.
It enables us to obtain a single predicted value of the spillover effect caused by the stimulus.
However, it is obvious that we cannot obtain the same result as that of the prediction,
even if we use exactly the same volume as that used in the calculation.
Nevertherless, we normally expect that the result should be substantially around the prediction.
This concern, that, whether the expectation is correct is the main topic of this study.



If the size of the spillover effect is around the average of the prediction,
it should be true that the propagation is never amplified or reduced through firm networks.
However, Gabaix showed that if the firm size distribution is fat-tailed, the hypothesis breaks down \cite{Gabaix11}.
In addition, Acemoglu et al. pointed out that microeconomic shocks may lead to aggregate fluctuations
in the presence of intersectoral input-output linkages \cite{Acemoglu12}.
These studies suggest that the stimulus and spillover effect
do not result in the proximity of the prediction.
In other words, normal distribution is usually assumed, but this assumption is doubtful.




This study reveals how demand shocks from outside cause the spillover effect.
We use a micro model invented by Bak et al. \cite{Bak93} and
employ observed data.
We clear up the following points:
(1) the diversity of the spillover effect. (it must depend on industries in which shocks are given.)
and (2) the extent of getting involved in the spillover effect. (it must also depend on industries.)


The remainder of this paper is organized as follows.
In Section 2, we introduce the dataset.
Section 3 describes the methodologies that we utilize in analyses.
Section 4 presents the results.
Finally, Section 5 concludes.

\section{Data}

We use datasets, TSR Company Information Database and TSR Company Linkage Database,
collected by Tokyo Shoko Research (TSR),
one of the major corporate research companies in Japan.
The datasets are provided by the Research Institute of Economy, Trade and Industry (RIETI).
In particular, we use the dataset collected in 2012.
The TSR data contain a wide range of firm information.
As necessary information for our study,
we use identification, capital, industry type, suppliers and clients.
We construct an entire network of firms based on suppliers and clients.
Note that there are up to 24 suppliers and up to 24 clients for each firm in the data.
It may be considered that the constraint limits the number of links for each node.
However, a node can be suppliers of other nodes without limitation,
as long as those clients designate the node as a supplier,
and vice versa.
Therefore, the numbers of suppliers or clients are not limited to 24.
The number of firms, that is, nodes, is 1,109,549.
The number of supplier-client ties, that is, links, is 5,106,081.
This network has direction and the direction is important in our study.

We split firms based on industries.
The industries are classified by the Japan Standard Industrial Classification \cite{Ministry13}.
We mainly use the division levels that have 20 classifications.
However, we make alterations to the classification.
Since the classifications ``S: Government, except elsewhere classified'' and ``T: Industries unable to classify''
are less important in our study, we omit them.
In addition, we separate ``I: Whole sale and retail trade'' into wholesale and retail.
The difference of the divisions is not negligible in our study because shocks from outside, such as fiscal policies,
often occur in retail.
Therefore, the division level in our study after alterations shows 19 industries.

Moreover, we use three industries at the group level to compare the effects of some Japanese fiscal policies.
The groups are
``5911:  New motor vehicle stores,''
``5931:  Electrical appliance stores, except secondhand goods,''
and ``6821:  Real estate agents and brokers.''
The difference between the division and group levels is clear in the later industries and there is no concern
about confusing the two levels.

We use an input-output table to compare the prediction of the spillover-effect size
between a micro model and the table.
As the closest table in time, we use the 2011 updated input-output table \cite{METI11}.

\begin{figure}[h]
\begin{center}
\includegraphics[width=15cm]{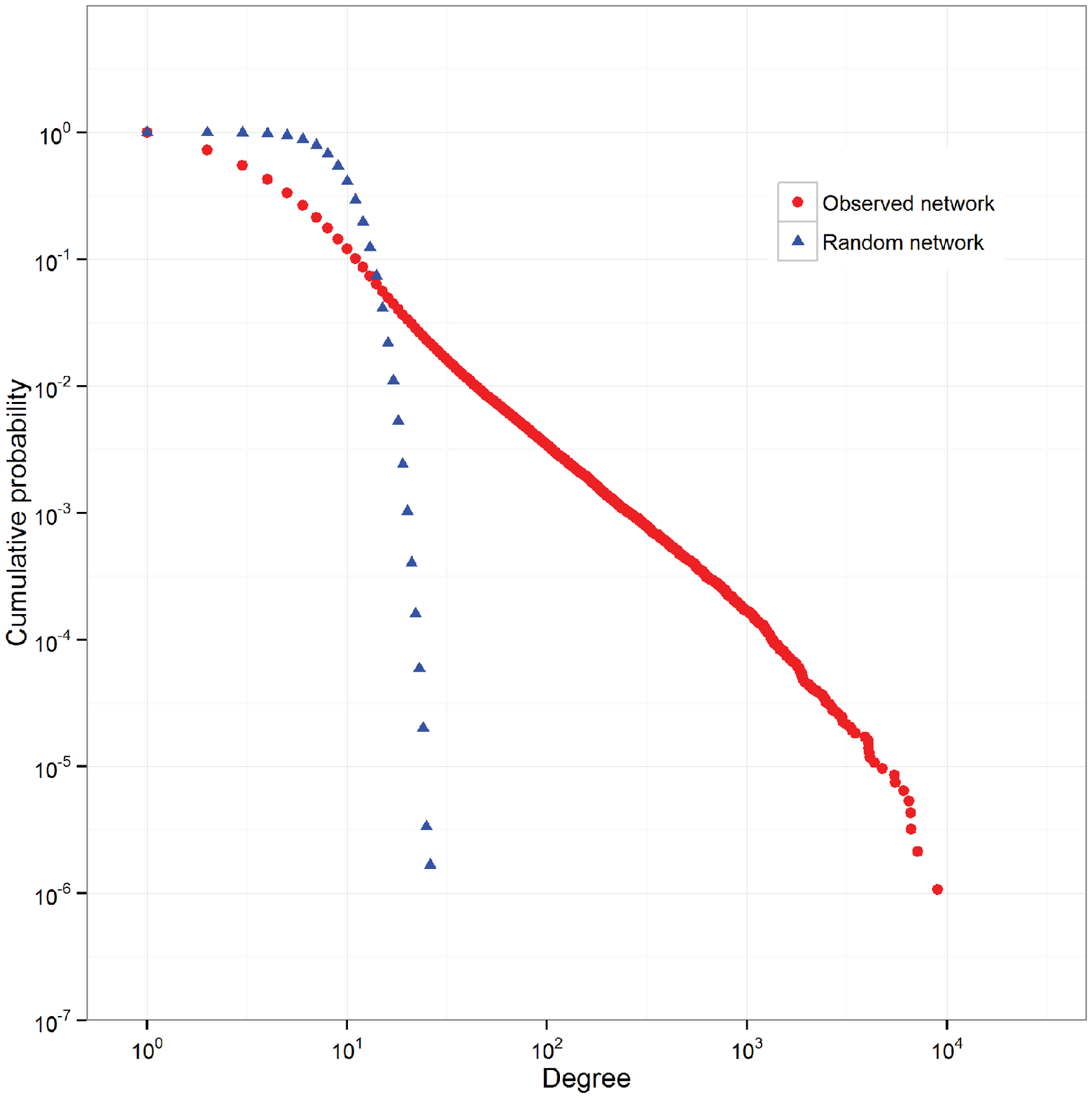}
\caption{Comparison of degree distributions between the random network and observed network:
The horizontal axis shows degree and the vertical axis shows cumulative probability.
The blue plots indicate the random network.
The red ones indicate the observed network.}
\label{fig:degreeRandOb}
\end{center}
\end{figure}

Figure \ref{fig:degreeRandOb} shows the degree distribution of the observed network.
The red plots are the distribution of the observed network.
An important point is that the distribution is fat-tailed,
which means the distribution does not decay super-linearly.
It seems that we can fit plots to a line $P\propto k^{-\lambda}$, where $P$ is the cumulative probability,
$k$ is the degree, and $\lambda$ is a positive constant.
If the degree distribution is the normal distribution,
the plot is shaped as the blue plots in Figure \ref{fig:degreeRandOb}.
Since normal distribution exponentially decays,
we can observe the blue plots decrease super-linearly on the log-log plot.
How to create the random network is explained in Section \ref{cha:results}.
The reason we should compare the observed network with the random network is that
the random network creates aggregate fluctuation that decays exponentially, as we show in Section \ref{cha:results}.
In other words, it tells us that if the random network is the network in the real economy,
there is no fat-tailed aggregate fluctuation.
However, the observed network is not the random network as we see here.

\begin{center}
[Figure \ref{fig:degreeRandOb} here]
\end{center}

If a probability distribution or a cumulative probability distribution
can be fitted to a line, it is said that the distribution is a power-law distribution.
A network with a power-law distribution is often called a scale-free network.
It has been pointed out that the power-law or scale-free nature of networks is a determinant
of fat-tailed aggregate fluctuations \cite{Gabaix11}.
Since the observed network is the scale-free network,
we expect that the aggregate fluctuations of the network are fat-tailed.

\section{Methodology}

\label{cha:Methodology}

We use a modified model \cite{Iino09} based on a production model \cite{Bak93}.
The modified model enables us to conduct micro-level simulations and investigate the characteristics of aggregate fluctuations.
The model of production and inventory was
originally invented by Bak et al. \cite{Bak93}.
The model assumes that
firms connect with each other on supply chains.
Each firm has some amount of inventory.
When a firm receives orders from clients, it supplies intermediate goods or services to clients.
If the firm does not have enough inventory, it sends orders to suppliers.
Therefore, cascades of orders and production sometimes occur.
The size of the cascades can be defined by the total extent of production due to activated firms. 
Bak et al. showed that the distribution of the cascade size follows the power law.
This result underlies recent network-based studies related to aggregate fluctuations \cite{Gabaix11,Acemoglu12}.
That is, the cascade reaction can be understood as aggregate fluctuations.
Here, for brevity, we call the cascade reaction an avalanche.



The result obtained by Bak et al. 
strongly depends on the regularity of supply chain network.
A node has two suppliers and two clients in the regular network,
except the nodes in the top and bottom layers.
As we have already shown that the real supply-chain network is
not a regular network but a scale-free network,
the assumption is too strong to apply the model to the real supply-chain network.
To mitigate the limitation of the regular network,
Iino and Iyetomi generalized the model so that a node has arbitrary numbers of in-degree or out-degree
and analyzed the nature of the generalized production model \cite{Iino09}.
We employ their generalized model with a minor modification.

\begin{figure}[h]
\begin{center}
\includegraphics[width=15cm]{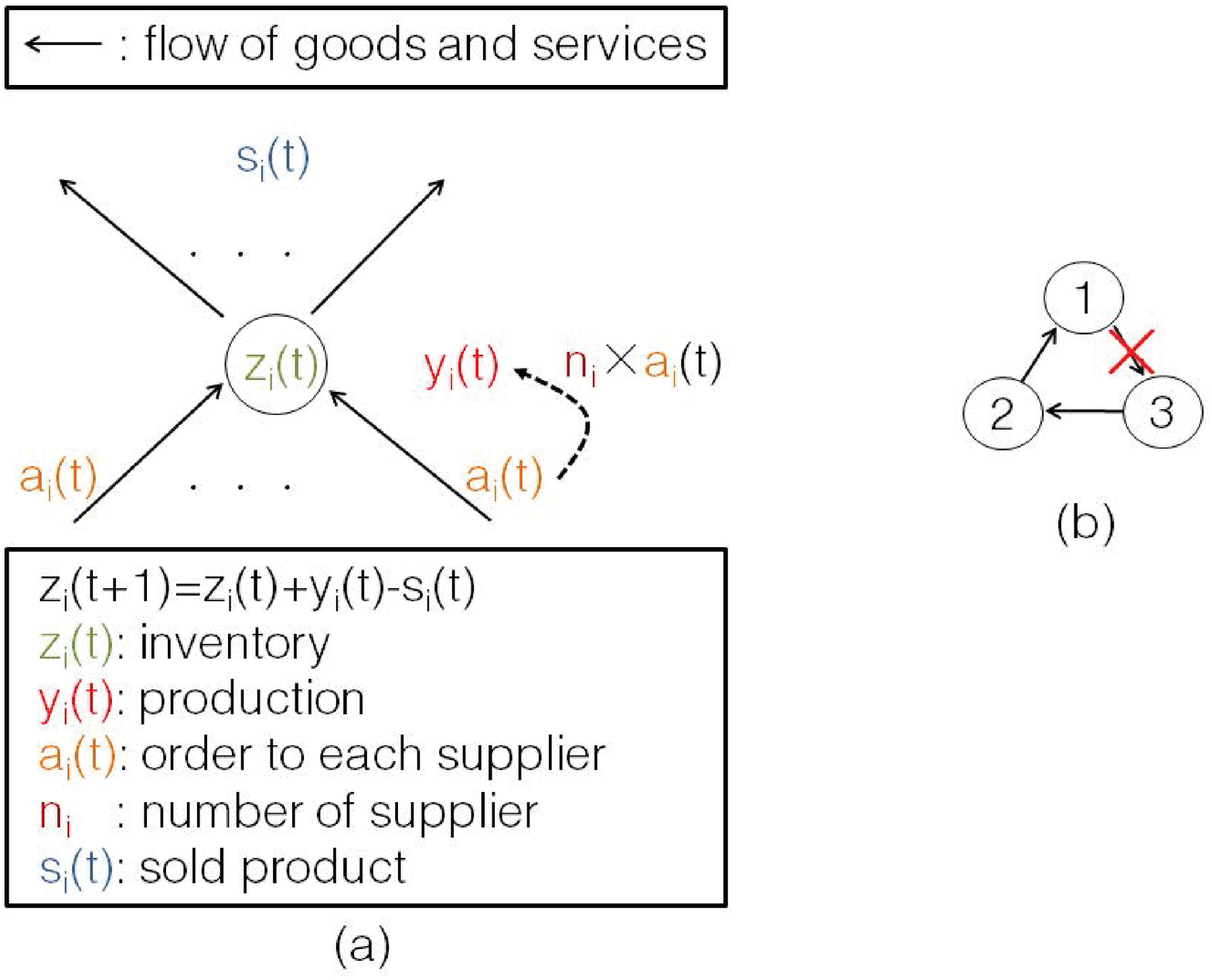}
\caption{Generalized production and inventory model:
(a) Scheme of the inventory renewal. Arrows show the flows of products.
Therefore, orders of supplies are opposite.
(b) Example of loop avoidance assumption.
There are three firms 1, 2, and 3.
Firm 1 asks firm 2 to supply products (given firm 1's inventory is not enough for demand.)
Firm 2 asks firm 3 to supply products.
However, firm 3 does not take supply from firm 1 and the link is ignored.
}
\label{fig:schemeProductionModel}
\end{center}
\end{figure}

Here, we describe the model used in our analyses.
For every time step $t$ and every firm $i$,
a new amount of inventory is decided based on the following equation.
\[z_{i}(t+1)=z_{i}(t)-s_{i}(t)+y_{i}(t),\]
where $z_{i}(t)$ is the amount of inventory of firm $i$ at time $t$,
$s_{i}(t)$ is the amount of orders received by firm $i$ at time $t$,
and $y_{i}(t)$ is the amount of production conducted by firm $i$ at time $t$.
The equation renews the inventory and is depicted in Figure \ref{fig:schemeProductionModel}(a).
We assume that (1) the firm equally sends orders to its suppliers,
(2) each firm produces one unit of production from one unit of material that it obtains,
and (3) a firm produces the minimum goods necessary to meet requests from its consumers.
Assumptions (1) and (2) simply result in a production feature
in which $y_{i}(t)$ is a multiple of $n_{i}$, where $n_{i}$ is the number of suppliers for the firm $i$.
In addition, the assumption (3) results in $z_{i}(t)\leq n_{i}$.
Based on the inventory renewal equation and the assumptions,
the amount of production $y_{i}(t)$ is given by
\begin{eqnarray}
\label{eqn:production}
y_{i}(t)=\left\{ \begin{array}{ll}
0 & (s_{i}(t)\leq z_{i}(t))\\
n_{i} & (z_{i}(t)<s_{i}(t)\leq z_{i}(t)+n_{i}) \\
\vdots & \vdots \\
a_{i}(t)\cdot n_{i} & (z_{i}(t)+(a_{i}(t)-1)n_{i}<s_{i}(t)\leq z_{i}(t)+a_{i}(t)n_{i}),\\
\end{array} \right.
\end{eqnarray}
where $a_{i}(t)$ is the number of orders that firm $i$ places with each supplier.
$a_{i}(t)$ is calculated by a ceiling function
\[a_{i}(t)=\ceil[\Bigg]{\frac{s_{i}(t)-z_{i}(t)}{n_{i}}}.\]
Since the quantity of the received orders $s_{i}(t)$ is the sum of orders to firm $i$,
\[s_{i}(t)=\sum_{j}{a_{j}(t)},\]
where $j$ is one of clients of firm $i$.
If a firm does not have a client and the firm needs to produce,
the firm regarded as belonging to primary industry and is assumed able to produce an arbitrary amount of production.

\begin{center}
[Figure \ref{fig:schemeProductionModel} here]
\end{center}

The first orders are placed from outside.
Depending on the analyses, a firm is selected from all firms or firms in a specific industry to place an order.
The selection is uniformly random.

Two firms may mutually supply
or a long step of supply chains may form a cycle.
It is possible that firms on a loop indefinitely produce goods or services,
although this never occurs in the real economy.
Iino and Iyetomi assume that firms have randomly assigned potential values,
which is an analogy of electrostatics.
A firm with more potential than another can supply, which is similar to water flow.
Although this assumption helps avoid a loop and is useful for analyzing the nature of randomly created networks,
it is not particularly clear how to assign a value to each firm.
Here, we make a simple assumption.
A firm that has already supplied products, that is, a firm that is already in a propagation process,
is ignored as a supplier.
Figure \ref{fig:schemeProductionModel}(b) shows an example of a loop with three firms.
Firm 3 ignores firm 1 when it needs production.
More precisely, the supply link from firm 1 to firm 3 is tentatively ignored.

Since the observed data include all industries,
it may be considered that it is unnatural to contemplate inventory for service industries.
This is because 
it is not understandable to consider inventory for intangible products, such as insurance or healthcare.
However, service industries {\it do} have inventory.
That is, if any service is ready to be used, it should be considered as inventory.
For example, a vacant hotel room ready for use incurs cost.
Therefore, we can discuss all industries that deal with tangible or intangible products
in the same network.

\section{Results}

\label{cha:results}

In this section,
we show the fat-tailed nature of avalanches
and their diversity over industries.
We start with the results of avalanches
comparing a random network and the observed network.
In a random network, every pair of nodes is connected according to constant probability $p$.
The expected number of links of the random network with $p$ is
$\left( \begin{array}{c}
N \\
2
\end{array} \right) p$
, where $N$ is the number of nodes.
The random network is created so that the network has the same number of nodes and the same number of links
as the observed network.
The observed network has 1,109,549 nodes and 5,106,081 links.
Therefore, we set $p$ as approximately $8.30\times 10^{-6}$.
As a result, we obtain a random network with 1,109,549 nodes and 5,366,223 links.
This accordance is necessary to compare the two networks.

The experiments proceed as follows for both networks.
(1) At time $t$, a firm is randomly chosen from all the firms.
(2) A chosen firm sells a unit of product.
(3) An avalanche is calculated.
(4) Repeat (1)--(3) for 1 billion times ($t$ proceeds from 1 to 1 billion.)
For an avalanche size, that is, aggregated production, we obtain
\[Y(t)=\sum_{i}y_{i}(t),\]
for every time step $t$.

\begin{figure}[h]
\begin{center}
\includegraphics[width=15cm]{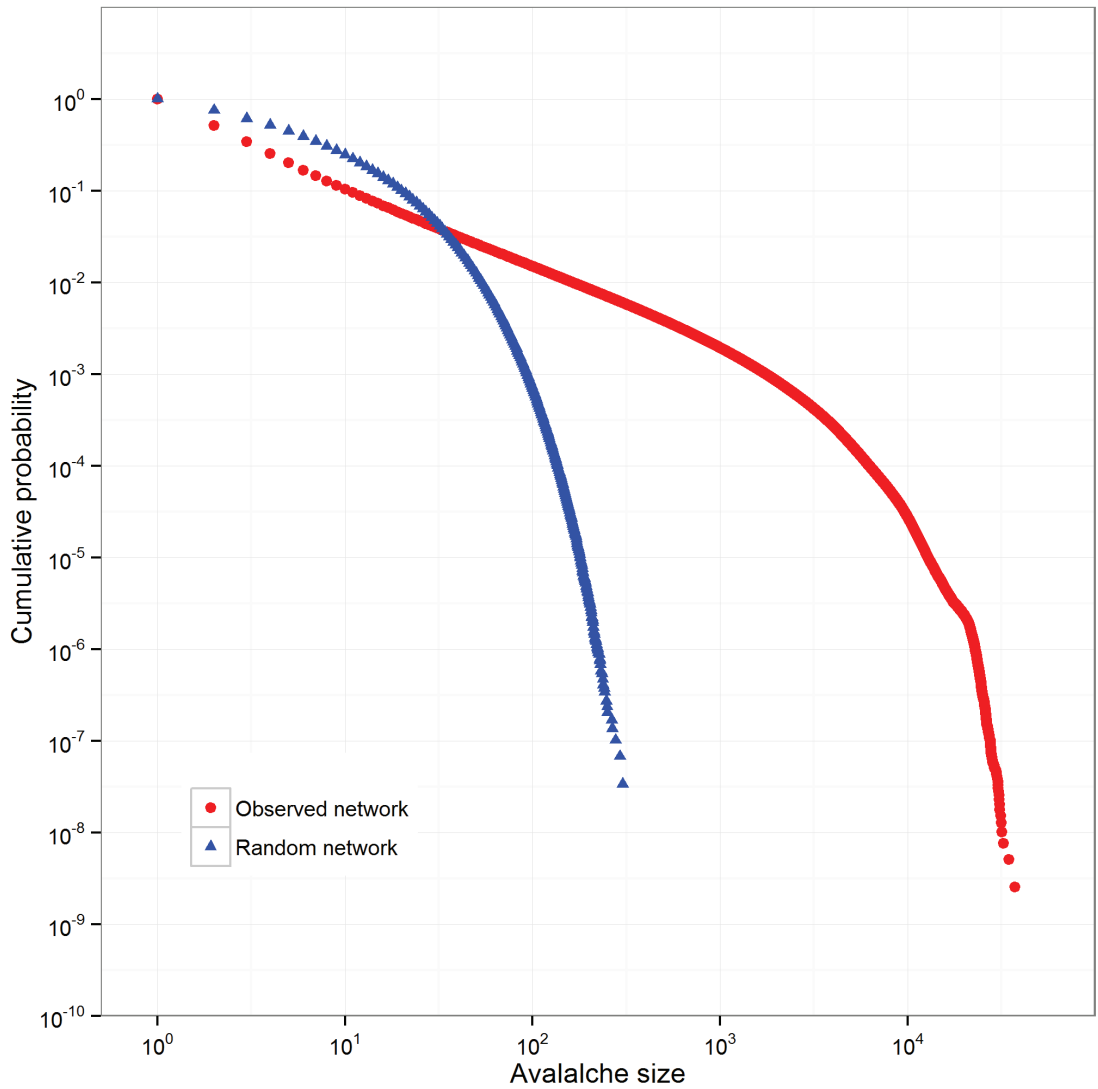}
\caption{Comparison of avalanche size distribution between the random network and the observed network:
The horizontal axis shows sizes of avalanches caused by demand.
The vertical axis shows the cumulative probability.
The blue plots indicate the random network.
The red ones indicate the observed network. Although there are size zero avalanches,
we ignore them
because the main aim of the figure is to show the shapes of the tails and zero cannot be included into log plots.}
\label{fig:compAva}
\end{center}
\end{figure}

Figure \ref{fig:compAva} shows the avalanche sizes for the two networks.
The red plots are for the observed network and the blue plots are for the random network.
The random network obviously decays fast and it seems that the distribution cannot be fitted to a line.
On the other hand,
the observed network has a part that can be fitted to a line.
This result, that a scale-free network has a fat-tailed avalanche size,
has already been shown analytically \cite{Iino09}
and has been proved partly under some constraints \cite{Zachariou15}.

\begin{center}
[Figure \ref{fig:compAva} here]
\end{center}

The results here tell us that uniformly random stimuli cause scale-free avalanche on the real network.
That is, the average avalanche size is not a representative value of it.
The input-output table analysis results in a single value of prediction,
which means some representative value is used to predict the spillover effect.
However, the power-law distribution does not have a typical scale.
It seems that careful analyses are required by the input-output table.

\begin{figure}[h]
\begin{center}
\includegraphics[width=15cm]{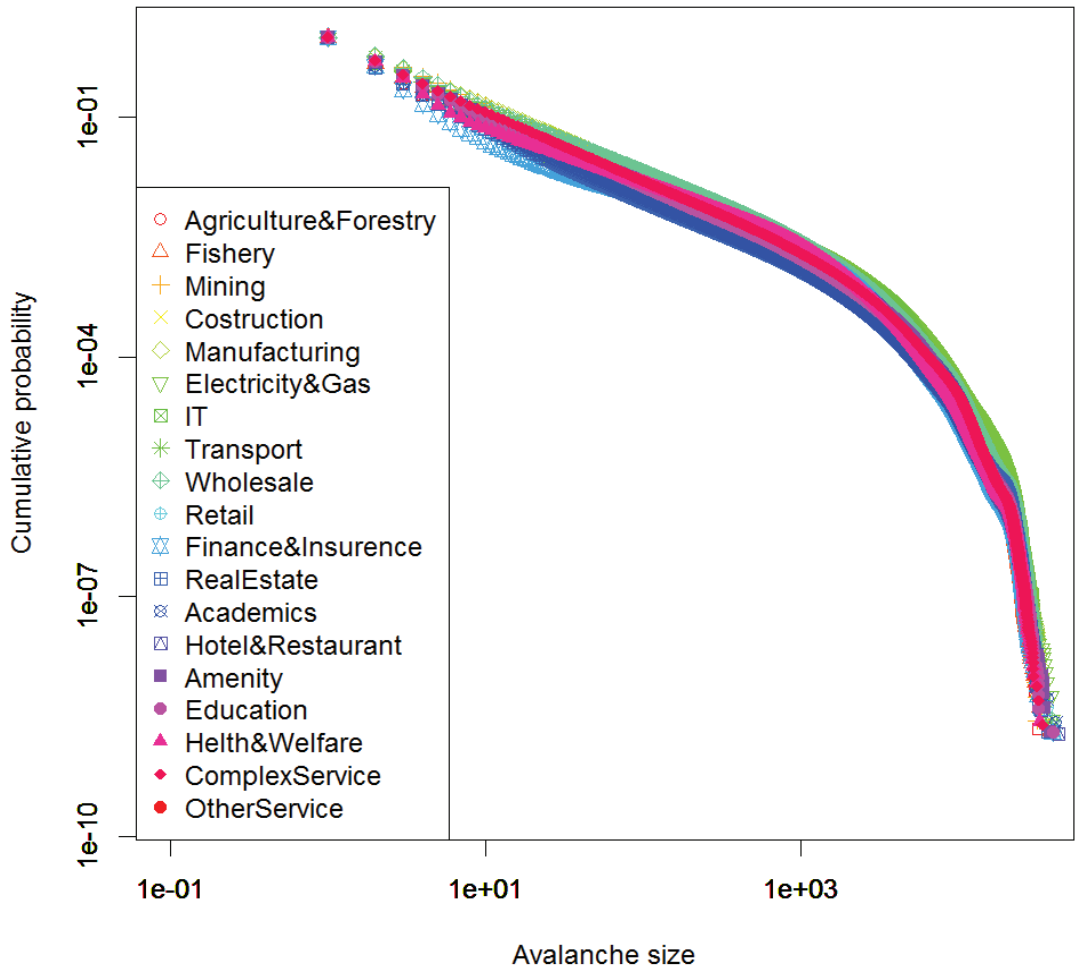}
\caption{Comparison of avalanche size distribution between industries:
The horizontal axis shows sizes of avalanches caused by demand.
There is repeated demand for a firm that
is chosen randomly from the industry.
The vertical axis shows the cumulative probability.
Although there are size zero avalanches,
we omit them
because the aim of the figure is to show the shapes of the tails and zero cannot be included in log plots.}
\label{fig:avaIndStartDist}
\end{center}
\end{figure}

From the viewpoint of fiscal policy,
it is important to know 
how stimulus received by different industries causes differences.
We conduct the experiments with a few changes to the previous experiments.
A firm is randomly chosen from an industry.
The industry is fixed through an experiment.
In every experiment, 1 billion instances of demand are given.
The experiments are conducted for the 19 industry divisions.
Figure \ref{fig:avaIndStartDist} shows the distributions of avalanche sizes.
While we had expected
that the distributions of avalanche sizes would have different shapes,
this turns out to be untrue.
As can be observed, there is no apparent difference in shapes.

\begin{center}
[Figure \ref{fig:avaIndStartDist} here]
\end{center}

\begin{figure}[h]
\begin{center}
\includegraphics[width=15cm]{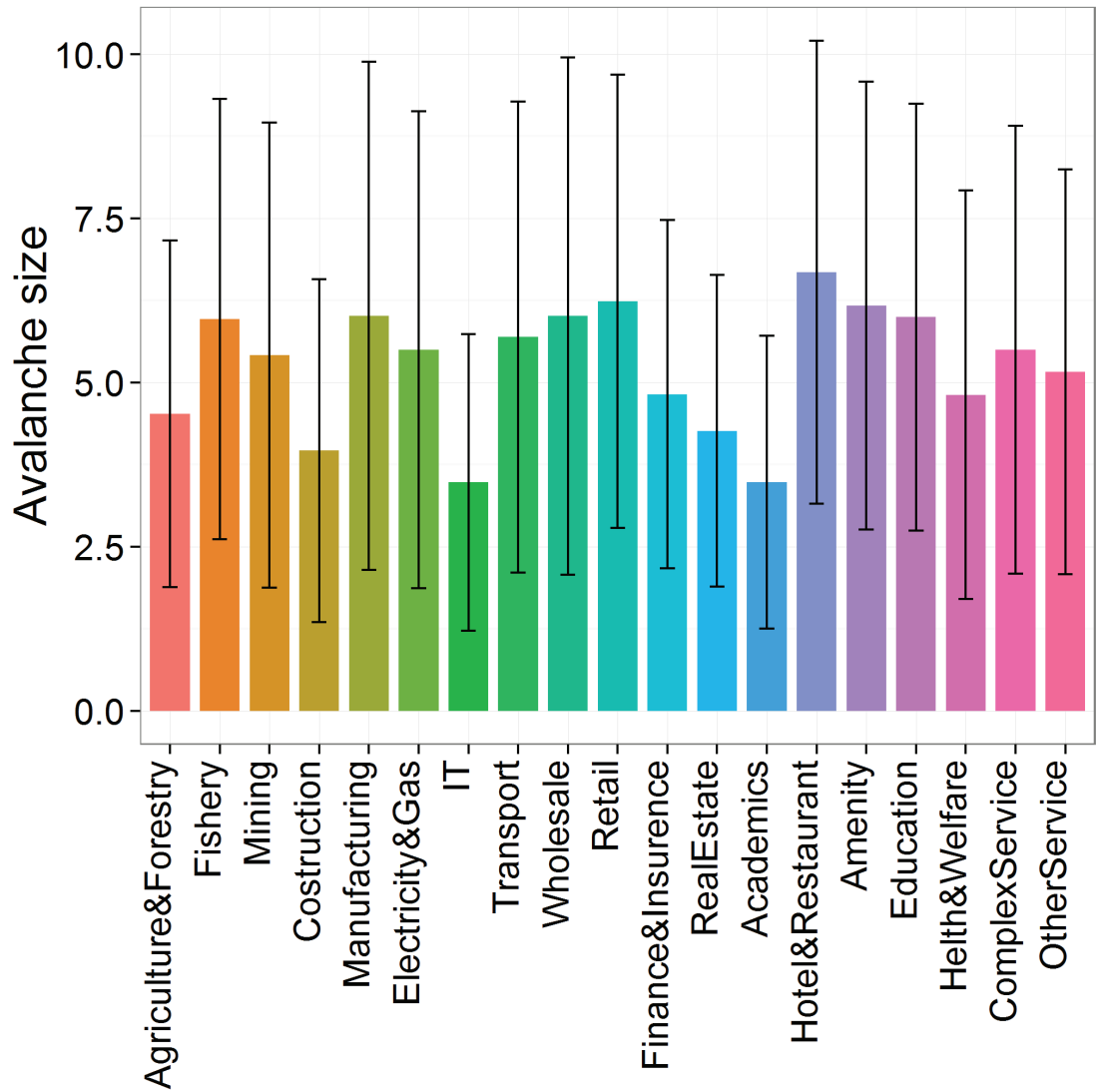}
\caption{Comparison of average avalanche size starting from specific industry:
The horizontal axis lists the industries. 
There is repeated demand for a firm that
is chosen randomly from the industry.
The vertical axis shows the average of avalanches.
The error bars show standard deviations.
}
\label{fig:avaIndStart}
\end{center}
\end{figure}

Figure \ref{fig:avaIndStart} shows the mean sizes of avalanches.
However, as mentioned in the last paragraph,
the size distribution is fat-tailed and therefore, the average is not a representative of the avalanche size.
The error bars in Figure \ref{fig:avaIndStart} show standard deviations.
Since we already observed that the avalanche size has the power law distribution in Figures \ref{fig:compAva} and \ref{fig:avaIndStartDist},
we know that standard deviations or variances are large.
We do not conduct the statistical test for the difference of the average
because the 1 billion samples cause
small standard errors and always show significance of the difference.
Therefore, the test is pointless.
Instead, it is important that standard deviations are overlapping.
It seems that we cannot strongly expect that the spillover effect started from a specific industry
that is certainly superior or inferior to other industries.

\begin{center}
[Figure \ref{fig:avaIndStart} here]
\end{center}

In Figure \ref{fig:avaIndStart},
the order of the industries that are aligned in the horizontal axis roughly shows the advancement from the primary industries.
Since an advanced industry, such as manufacturing or services, is in the downstream of supply chains and has long chains from the primary industries,
it may be considered that the distance to the primary industries has a positive correlation coefficient with the avalanche size of an industry.
However, we do not observe such a relationship,
which can be attributed to the network structure.
Scale-free network is a small-world network \cite{Watts99}.
A firm has a short path to a hub.
Since the hub yields a large $y_{i}(t)$,
the magnitude of avalanches seems to be dominated by this mechanism.

It may be argued that 
the production model of this study is only a model
and it is doubtful how much it can explain the actual economy.
Therefore, we simply compare the size of avalanches and
the inverse matrix of the input-output table \cite{METI11}.
The Pearson's correlation coefficient is 0.28.
If we consider that the production model does not include the data of trade volume at all,
it can be said that the coefficient is surprisingly large.
It should be noted that the production model can simulate the variances that cannot be obtained from the input-output table.

\begin{figure}[h]
\begin{center}
\includegraphics[width=8cm]{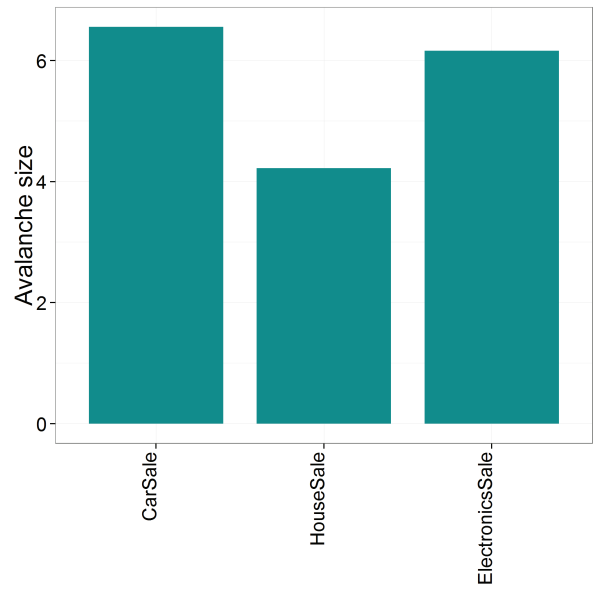}
\caption{Simulation for industries to compare fiscal policies:
The horizontal axis lists the industries.
They correspond to the target industries of past Japanese fiscal policies:
eco-vehicle tax breaks, eco-point system for housing, and eco-point system for home electronics.
The vertical axis shows the average of avalanches.
There is demand for a firm in each industry.
}
\label{fig:avaPolicy}
\end{center}
\end{figure}

We conduct further experiments that start from the group level of industries
so that we can compare the effect of actual policies.
The results of Figure \ref{fig:avaPolicy} are based on the division level industries.
Those industries are
``5911:  New motor vehicle stores'' (CarSale),
``5931:  Electrical appliance stores, except secondhand goods'' (ElectronicsSale),
and ``6821:  Real estate agents and brokers'' (HouseSale), which are at the group level.
They correspond to the target industries of past Japanese fiscal policies:
eco-vehicle tax breaks, eco-point system for housing system, and eco-point system for home electronics.
The setup of experiments is the same as that for the previous experiments.

\begin{center}
[Figure \ref{fig:avaPolicy} here]
\end{center}


Since the government publishes the actual size of budgets and some institutes publish the estimated economic results,
we can validate the predictability of the model.
The government lost tax revenue corresponding to 241.0 billion Japanese yen
(1.98 billion US dollars at an assumed exchange rate of 122 Japanese yen to 1 US dollar) in 2009
for the eco-vehicle tax breaks \cite{Ministry15}.
The economic result was estimated about 2,937.53 billion Japanese yen (24.08 billion US dollars) \cite{Shirai10,JapanAutomobileDelersAssociation15}.
However, this estimation does not include indirect effects of other industries.
The leverage of the return to the investment is 12.18.
The government spent 494.8 billion Japanese yen (4.06 billion US dollars)
for the eco-point system for housing in total.
The economic result was estimated at about 414.0 billion Japanese yen (3.39 billion US dollars) \cite{MizuhoResearchInstitute12}.
The leverage is 0.84.
The government spent 692.9 billion Japanese yen (5.68 billion US dollars)
for the eco-point system for home electronics in total.
The economic result was estimated at about 5 trillion Japanese yen (40.98 billion US dollars) \cite{BoardOfAuditJapan12}.
The leverage is 72.16.
Although the economic result and leverage are just estimations in the abovementioned studies,
the eco-point system for housing is apparently small, as we find in our experiments.
Three samples are small but we find no apparent contradiction in the results.


We observe the distributions of avalanche sizes that are stated from specific industries in Figure \ref{fig:avaIndStartDist}
and the shapes of the tails are similar.
This can be interpreted as
a certain supply chain always being used in those large avalanches
and the supply chain may lie on particular industries.
To examine the hypothesis,
we obtain a different measure from the first experiments.
The measure is how often firms in an industry become involved in avalanches.
Note that we examine the avalanche sizes that start from specific industries thus far and a firm is randomly selected from all firms.

\begin{figure}[h]
\begin{center}
\includegraphics[width=15cm]{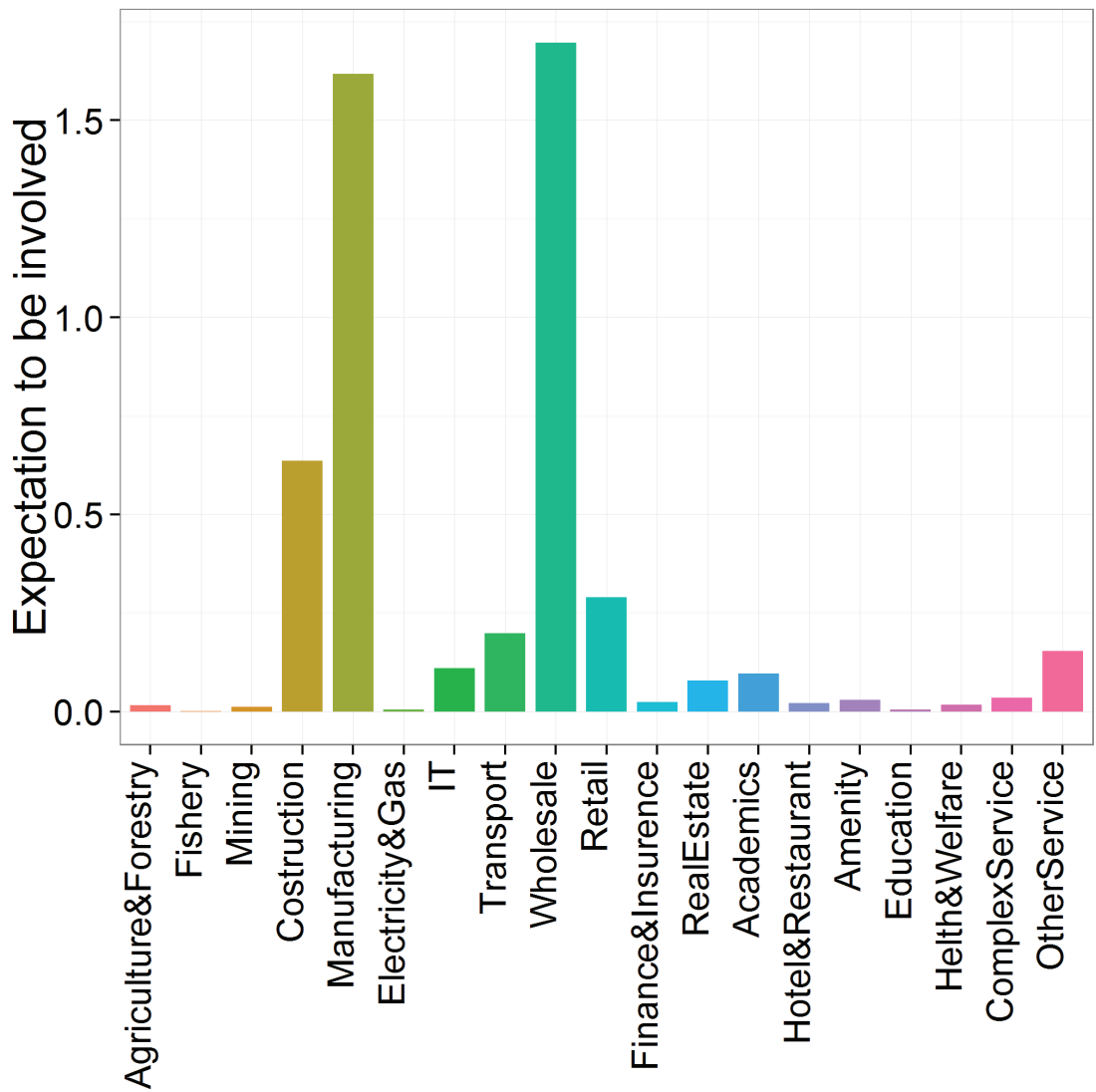}
\caption{
Difference of expectations for being involved in avalanches:
The horizontal axis lists the industries.
The vertical axis shows the expectation to be involved in avalanches per instance of demand.
The firm with demand is chosen randomly from all the firms.
The expectation in each industry is averaged over the firms in each industry.
}
\label{fig:expInvolved}
\end{center}
\end{figure}

Surprisingly, as is observed in Figure \ref{fig:expInvolved},
the extent to which a firm becomes involved in avalanches is sharply different.
The sharpness is totally different from that observed in Figure \ref{fig:avaIndStart}.
Wholesale and manufacturing are distinctly large and construction can be included in the largest group.
This result means that firms in those industries are apparently always involved in large avalanches
that start from any industry.

\begin{center}
[Figure \ref{fig:expInvolved} here]
\end{center}

\section{Conclusion}

This paper analyzed the cause of the diversity of the spillover effect.
We used observed data of transactions in Japan.
For the data, we employed the production model.
As a result, we confirmed the size of the spillover effect triggered by demand follows the power law.
Therefore, the normal distribution, which is usually expected in analyses of input-output tables,
cannot be a reliable assumption.
Although we did not use the volume of trade, the results of the simulations show significant correlation coefficients.
Moreover, the simulated avalanche sizes for the policies actually conducted correspond to the estimation given
by the ex-post evaluations of the policies.
In addition, industries have diverse potential to become involved in avalanches.

\bibliographystyle{unsrt}       
\bibliography{paper}

\clearpage

\clearpage

\clearpage

\clearpage

\clearpage

\clearpage

\clearpage

\end{document}